\documentclass[amsmath,amssymb,aps,prd,nofootinbib,12pt]{revtex4-1}

\usepackage{graphicx}
\usepackage{hyperref}
\usepackage{xcolor}
\usepackage{stackengine}

\begin{document}

\title{Thermodynamics of $f(R)$ Theories}
\author{S.E. Jorás}
\affiliation{
 Instituto de F\'\i sica, Universidade Federal do Rio de Janeiro,\\
 CEP 21941-972 Rio de Janeiro, RJ, Brazil 
}
\begin{abstract}
This series of three lectures was presented at ``Escola de Cosmologia e Gravitação" \url{https://cosmosecontexto.org.br/ecg-inscricoes} and webcast at \url{https://shorturl.at/2ZSI7} --- in Portuguese, but slides in English.

We will go through a brief review on $f(R)$ theories (in the metric approach) and the usual requirements for successful modifications of General Relativity. Then we will open a large parenthesis to talk about a non-standard approach to Phase Transitions: the Catastrophe Theory. Finally, we will connect the previous lectures to introduce a new Thermodynamic interpretation of $f(R)$ theories.   
\end{abstract}

\maketitle

This a very short review on $f(R)$ theory in the metric approach along with a recently developed Thermodynamics interpretation. The main goal here is to detail such new approach, making an analogy to Thermodynamics --- which is written in terms of Catastrophe Theory. The latter deals with the coalescence and birth of extrema in an ordinary (polynomial) function as a given set of free parameters are varied. We will see that is precisely what happens at a first-order phase transition and, therefore, it describes exactly the exit from the inflationary regime in the early universe. 

Long reviews on $f(R)$ can be found in Refs.~\cite{Sotiriou_2010, sotiriou2007modified,2010LRR....13....3D} as well as Chap.~9 of Ref.~\cite{2015daen.book.....A}.
A modern approach to phase transitions can be found in Ref.~\cite{1985tait.book.....C}. Reviews on Catastrophes are available in Refs.~\cite{gilmore-1981a, poston1996catastrophe,saunders_1980}

\section{$f(R)$ Theories}

\subsection{Motivation}

The motivations for modifying such a successful theory as General Relativity (GR, from now on) can be divided in two (non-exclusive) categories: theoretical and experimental evidences.
The former usually relies on arguments from either the absence of renormalizability, or the first corrections from a possible theory of quantum gravity (whatever that might be), or just because that is the way to better appreciate the power and beauty of GR\footnote{The latter is my personal choice, I admit.}. The latter case --- experimental evidence --- is also compelling. We believe the universe experienced, in its early stages,  a short but tremendously strong phase of almost-exponential expansion --- inflation --- and/or a bounce (a contraction phase followed by expansion), besides the current mild accelerated expansion. The latter may be ``explained" by an {\it ad-hoc} Cosmological Constant $\Lambda$, which must be seen either as a new constant of nature (such as Newton's Gravitational Constant $G_N$ itself, for instance) or as a consequence from a more fundamental principle (such as a suitably renormalized ``vacuum energy"). Even so, that is not the case for the primordial accelerated phase, since a simple Cosmological Constant $\Lambda$ wouldn't have allowed an exit (graceful or not) from inflation. The universe would not have been through decelerated expansion (radiation- and matter-dominated) phases. In either case, the standard replacement for $\Lambda$ is a scalar field, with, truth to be told, not-so-standard potential end/or kinetic terms (k-essence) and couplings. 

I should say that some authors do investigate if modifications of GR can explain the rotation curves of galaxies (see, for instance, works on MOND \cite{1983ApJ...270..371M,Marra_2020}), but we will not pursue this goal here. 

Either way, we are currently at a crossroads in GR, just like Newton's Gravitation (NG) was in the XIX century, at two similar situations, with opposite results. In one of them, we relied on NG to explain small disturbances on the orbit of Uranus, which ultimately led to the discovery of Neptune. On the other hand, the explanation of Mercury's orbit could {\it not} be explained by an inner planet called Vulcan (simply because it was never found at the ``right" position) and, ultimately, GR extended NG. Now, like then, we have to decide if we should change the theory  or add an unknown component of the universe (a new ``planet'', particle or   scalar field, perhaps?) but keep the theory (GR).

\subsection{Modified Einstein Equations}

Here, we will focus on the so-called $f(R)$ theory in the metric frame, where the metric $g_{\mu\nu}$ is the one independent variable. We just briefly mention the Palatini \cite{2011IJMPD..20..413O} and Metric-Affine \cite{2007AnPhy.322..935S,hernandezarboleda2023palatini} approaches, according to which the Affine Connection $\Gamma^\alpha_{\mu\nu}$ is an independent object as well. 

Of course, if $f(R)=R - 2\Lambda$, then we have GR with a standard Cosmological Constant. Therefore, we shall look for non-linear functions only. To be more accurate, we replace the standard Einstein-Hilbert Lagrangian (plus an extra term that describes the matter/radiation/field content of the universe)
\begin{equation}
{\cal L}_{GR} = \sqrt{-g} \frac{1}{2\kappa^2}(R - 2\Lambda) + {\cal L}_m
\label{EH}
\end{equation}
by 
\begin{equation}
{\cal L}_{GR} = \sqrt{-g} \frac{1}{2\kappa^2} [f(R) - 2\Lambda]+ {\cal L}_m,
\label{Lf}
\end{equation}
where $\kappa^2 \equiv 1/M_{pl}^2 \equiv 8\pi G$. 
The former yields the Einstein Equations
\begin{equation}
R_{\mu\nu} - \frac{1}{2} g_{\mu\nu} R + \Lambda g_{\mu\nu} = \kappa^2 T_{\mu\nu},
\end{equation}
while the latter yields
\begin{equation}
R_{\mu\nu} f' -\frac12  g_{\mu\nu} f + g_{\mu\nu}\, \Box f' - \nabla_{\mu}\nabla_{\nu} f' = \kappa^2  T_{\mu\nu},
\label{mee}
\end{equation}
which looks (and actually is) much more complicated! For starters, one notices that it is a fourth-order differential equation in the metric (we recall the reader that the Ricci scalar $R$ itself depends on second-order derivatives on $g_{\mu\nu}$). As such, it bears an extra degree of freedom, that will become even more explicit shortly.

For now, we can make Eq.~(\ref{mee}) more palatable if it is rewritten as
\begin{equation}
R_{\mu\nu} - \frac{1}{2} g_{\mu\nu} R  = \kappa^2 T_{\mu\nu}+ \Bigg[R_{\mu\nu} - \frac{1}{2} g_{\mu\nu} R -\bigg( R_{\mu\nu} f' -\frac12  g_{\mu\nu} f + g_{\mu\nu}\, \Box f' - \nabla_{\mu}\nabla_{\nu} f'\bigg)\Bigg].
\end{equation}
The term inside the square brackets on the right-hand side can be read as an effective energy-momentum tensor for the so-called ``curvature fluid" $\kappa^2 T_{\mu\nu}^{(c)}$ --- since it is pure geometry. Such form is more than adequate (as opposed to an alternate form that divides the previous equation by $f'$, that will not be adopted here) because the energy-momentum tensor of the standard components and its conservation does not need to change. Consequently, the curvature fluid is conserved by itself as well. 

One can further define the energy density $\rho_c$ and pressure $p_c$ for such fluid, as
\begin{align}
\kappa^2 \rho _{c} &\equiv \frac{1}{2}\left( f'R - f\right) - 3H\dot{f'} + 3H^{2}(1-f') \label{rhocurv}\\
\kappa^2 p_{c}  &\equiv \ddot{f'} + 2H\dot{f'} - (2\dot{H}+3H^{2})(1-f') + \frac{1}{2}(f-f' R) 
\label{pcurv}.
\end{align}
One can then proceed to investigate the system as if it were described by GR in the presence of two fluids: the curvature fluid defined above and the standard matter/radiation fluid. We point out that the curvature fluid does not need to satisfy the standard (strong, weak, null) energy conditions, since it is not an actual physical fluid.

\subsection{The Frames}

In this section, we will make it explicit the extra degree of freedom inherent to $f(R)$ theories. We start over from the Lagrangian (\ref{Lf}) and write it as
\begin{equation}
{\cal L}_g = \sqrt{-g} \frac{1}{2\kappa^2} \bigg[ \phi R(\phi) - W(\phi)\bigg] + {\cal L}_m
\end{equation}
where
\begin{align}
\phi & \equiv f'(R) \\
 W(\phi) &\equiv \phi R(\phi) - f[R(\phi)]
\end{align}
which is simply a Legendre Transform, that replaces $R$ by $\phi$ as the independent variable. For that, we require a definite sign for $f''\equiv d^2f/dR^2$ (usually taken positive), except, perhaps, at particular values of $R$.

We now perform a conformal transformation where the conformal factor is $\phi$ itself:
\begin{equation}
g_{\mu\nu} \rightarrow \tilde{g}_{\mu\nu} \equiv \phi \cdot g_{\mu\nu},
\label{conformal}
\end{equation}
and define a new field 
\begin{equation}
\tilde\phi \equiv 
 +\sqrt{\frac{3}{2\kappa^2}}\ln\phi,
\end{equation}
which requires that $f'\equiv \phi >0$, already assumed when we performed the conformal transformation (\ref{conformal}). Note that, since the field $\phi$ is dimensionless and $\kappa \equiv 1/M_{pl}$, then $\tilde \phi$ has the dimensions of $M_{pl}$.

Upon the definitions above, Eq.~(\ref{Lf}) is written as
\begin{equation}
\tilde {\cal L}_g = \sqrt{-\tilde g} \bigg[ \frac{1}{2\kappa^2}\tilde R - \frac{1}{2}\tilde g^{\alpha\beta} \partial_\alpha \tilde\phi \partial_\beta \tilde\phi -V(\tilde\phi)\bigg] + {\cal L}_m \bigg[\psi, \frac{1}{\phi} \tilde g_{\mu\nu}\bigg]
\label{Ltil}
\end{equation}
where
\begin{equation}
V(\tilde\phi) \equiv \frac{R(\phi) \phi - f[R(\phi)]}{2\kappa^2\phi^2}
\end{equation}
and the determinant $\tilde g$ and the Ricci scalar $\tilde R$ are calculated from the  new metric $\tilde g_{\mu\nu}$. 

The Lagrangian (\ref{Ltil}) is the one of GR with an extra scalar field $\tilde\phi$, which carries the extra degree of freedom, as previously announced. One should notice that the field $\psi$ (that represents matter/radiation) does not follow geodesics in the new frame, although its equation of motion should be expressed in terms of the new metric. Accordingly, it is not conserved if the covariant derivative is calculated in terms of $\tilde g_{\mu\nu}$, since this is supposed to hold in the previous metric only. One can also notice that $\psi$ is not minimally coupled to the metric $\tilde g_{\mu\nu}$ since $\phi$ shows up

This is the so-called Einstein frame. Its ``equivalence" to the Jordan frame (before the conformal transformation) is a topic of debate in the current literature, along with the question about which one is the ``physical" one (whatever that means). Here, we will rely on the conservation of the energy-momentum tensor of matter/radiation in the Jordan frame in order to define that all the observations are supposed to be done in that frame. One can, obviously, use the Einstein frame to make calculations, but, at the end of the day, one should go to the Jordan frame in order to make predictions on observable quantities. With that in mind, we point out that eventual divergences in the EF are not always bad, if they are mapped into finite quantities in the Jordan frame.

One final piece of information: the aforementioned ``equivalence" between the frames is robust when it comes to inflation. Indeed, it is possible to show \cite{1995PhRvD..52.4295K} that if the slow-roll parameters hold in one of the frames, then they also hold in the other one.

\subsection{Constraints}

One can always write Eq.~(\ref{Lf}) as 
\begin{equation}
{\cal L}_f = \sqrt{-g} \frac{1}{2\kappa^2} f(R) = \sqrt{-g}  \frac{1}{2\kappa^2} \bigg(R + \epsilon \Delta(R)\bigg) = \sqrt{-g}  \frac{1}{2\kappa^2} R \bigg(1 + \epsilon \frac{\Delta(R)}{R}\bigg),
\label{Delta}
\end{equation}
which defines $\Delta(R)$ and $\epsilon$, where the latter indicates the amplitude of the deviation $\Delta(R)$ from the Einstein-Hilbert Lagrangian ${\cal L}_{EH} \equiv R$.
The standard constraints on $f(R)$ theories are almost always listed then as \cite{PhysRevD.77.023503}
\begin{align}
~& \lim_{R\to\infty} \frac{\Delta}{R} =0\\
~& \phi \equiv f'\equiv 1 + \Delta' >0\\
~& f''\equiv \Delta '' >0.
\end{align}
Nevertheless, they should be taken with a grain of salt. Let us look at them, one by one, and see why.

The first one states that the deviations from GR should be small when the Ricci scalar is large, i.e, in the primordial universe. That relies on the fact that we indeed understand the evolution of the universe from nucleosynthesis on as one of the many success of GR. Nevertheless, this claim is poorly stated. The reason is threefold. First, one aims exactly at non-negligible deviations from GR when modified theories are applied to explain the primordial inflationary era. Secondly, even when the extra terms $\Delta(R)$ are negligible, the equations of motion (EoM) are still higher-order differential equations. As such, there is no guarantee that their solutions are going to be closer to the ones from GR. Actually, it can be easily seen that this is not the case if one adds higher-order terms at the EoM of {\it linear}  harmonic oscillator. (except with strong fine tuning of the initial conditions). An even simpler case (a homework) is to compare solutions of the equation $\epsilon \ddot x + \dot x = 0 $ in the limit $\epsilon\to 0$ to the ones from the ``original" equation $\dot x = 0 $. 
Obviously, in modified theories of gravity, the nonlinearities only make it worse.  Finally, the Ricci scalar is not necessarily large in the early universe --- we recall the reader that the trace of Eq.~(\ref{mee}) is now
\begin{equation}
3\square f_R(R) + Rf_R(R) - 2f(R) = \kappa^2 T 
\label{trace}
\end{equation}
(where $T$ is the trace of $T_{\mu\nu}$) which is not an algebraic relation anymore (as in GR). Therefore, one could still have a large $T$ (as it is expected in the early universe), but a small $R$.  

The second constraint usually implies that it is required for an attractive gravity. It relies on dividing Eq.~(\ref{mee}) by $f'$ and noticing that it would change the sign of the Gravitational coupling $G$. Nevertheless, that mapping would disrupt the very conservation of $T_{\mu\nu}$ and we will not follow it here. However, it is indeed necessary to perform the conformal transformation to the Einstein frame (which is desired, but it is far from being an actual constraint). 

Finally, the third constraint is well-posed. It can be shown that perturbations of the flat space (Minkowski) are stable if and only if $f''>0$. Here, we follow Ref.~\cite{Sotiriou_2010}. Let us expand Eq.~(\ref{trace}) around Minkwoski metric $\eta_{\mu\nu}$, i.e,
\begin{align}
g_{\mu\nu} &= \eta_{\mu\nu} + h_{\mu\nu}\\
R &= T + \delta R,
\end{align}
and assume Eq.~(\ref{Delta}), which yields
\begin{equation}
\ddot {\delta R} + \frac{1}{2\Delta''}\bigg( \frac{1}{\epsilon} - \Delta' \bigg) \delta R \approx 0.
\end{equation}
Since $\epsilon$ can be as small as needed, the term inside the parenthesis is always positive. The sign of the effective mass squared for $\delta R$ is the same of $\Delta"$, which should be positive for stable perturbations.
We will see further below that the same constraint can be obtained from different and independent calculations.

\subsection{Practical Applications}

\subsubsection{Cosmology}

When it comes to the background evolution of the universe, one should require that we should ``start" in a Radiation-Dominated (RD) universe, pass through a Matter-Dominated (MD) phase --- that should last long enough so that matter perturbations can grow --- followed by the current accelerated Dark-Energy Dominated (DED) phase. One could also add the inflationary era at the very beginning, but let us postpone that discussion for now, because it involves the (p)reheating process.

Using the language of dynamical systems, we should then require that the RD phase is {\it not} an attractor (otherwise, the universe would never have left it). Analogously, the MD phase has to be a saddle point (i.e, with at least one stable direction and another unstable one), so that the universe will be drawn towards it (after RD, along the stable direction) but move on to the DED phase (along the unstable direction). Usually, one requires that the last phase is an attractor (i.e, only negative eigenvalues) so that the initial conditions do not have to be fine tuned.  

Using that approach, Ref.~\cite{Amendola_2007} has determined the necessary conditions for any $f(R)$ so that the picture described above is fulfilled. The authors define the quantities
\begin{align}
m &\equiv \frac{R f''}{f'} \\
r &\equiv - \frac{R f'}{f},
\end{align}
which measure the deviation from GR --- for which $m=0$ and $r=-1$. 

The existence of a MD phase with the required constraints is translated to
\begin{equation}
\left\{
\begin{tabular}{l}
$m(r\approx -1) \approx 0^+$   ~ and\\
$\frac{dm}{dr}\big|_{r\approx -1}>-1$ ,
\end{tabular}
\right.
\end{equation}
while a de Sitter attractor requires either
\begin{align}
&\left\{
\begin{tabular}{l}
$m=-r-1$,\\
$\frac{\sqrt{3}-1}{2}<m<1$,\\
$\frac{dm}{dr}\big|_{r\approx -2}<-1$
\end{tabular}
\right.\\
&{\rm or} \\
& 0<m(r=-1) \leq 1.
\end{align}
That paper has shown that simple expressions for $f(R)$, such as simple power laws just will not work. 

The growth rate of perturbations can be affected, of course, by modifications on gravity. There is a couple of mainstream approached for this problem. One may, for instance, realize that there are two opposite regimes, known as weak- and strong-field limits. They correspond to situations where the chameleon effect (see next section) is in action or not, which can be modeled by an effective Newton's constant equal to $G_N$ (its standard value) or to $\frac{4}{3}G_N$ \cite{Herrera_2017}. 

A second approach is to model the value of $G_N$ as a function of wavenumber $k$ \cite{2008PhRvD..78b4015B}. Being dependent on scale, gravity will produce different effects at different distances from a central mass and, then, the so-called shell crossing \cite{2012PhRvD..85f3518B} --- a top-hat overdensity will then not retain its shape.

The final goal in both approaches is to obtain the number of observed objects in a given mass range, as a function of its redshift. That number can then be used to constraint the free parameters of your chosen model.

A short review can be found in Ref.~\cite{Batista_2021}.

\subsubsection{Relativistic Stars}

I briefly mention that, as a serious candidate for a full theory of gravity, a given $f(R)$ must be tested in different scenarios, such as, relativistic stars. 

Spherically-symmetric stars are the perfect equivalence to homogeneous cosmology, since they both have only one independent variable: in the former,  the radial coordinate $r$; in the latter, the time $t$. Both cover a large range of values of densities (from the core to the radial infinity, in stars) and have plenty of data available. 

The catch is the uncertainty with respect to the equation of state of the standard (sometimes not much so) matter that makes up the star. Using the aforementioned ``curvature fluid" analogy, there is clearly some degeneracy between the contribution from such a term and from a different matter composition. Nevertheless, one may still provide constraints on particular models, testing both the maximum mass (and minimal radius) and the stability of such configurations (usually against radial perturbations). Indeed, the latter are able to yield the strongest constraints on the free parameters of your theory \cite{2020JCAP...11..048P,2022JCAP...09..058P,2023arXiv230800203P}.

One caveat is the many non-equivalent definitions of mass \cite{2020PDU....2700411S}: the astrophysical --- the one measured by an observer at radial infinity, that can be calculated from an asymptotic expansion of the metric components, given by
\begin{equation}
M_{\rm astro} \equiv \lim_{r\to\infty}  \frac{r}{2G_N}\bigg( 1 - \frac{1}{g_{rr}}\bigg),
\end{equation}
or the plain integration of the energy density (with or without the proper-volume factor $\sqrt{-g}$) bounded by the star's surface at $r=r_*$
\begin{align}
M_\rho &\equiv 4\pi \int_0^{r_*} \rho(r) r^2 \, dr\\
M_{\rm prop} &\equiv 4\pi \int_0^{r_*} \rho(r) r^2 \, \sqrt{-g_{rr}} \, dr
\end{align}
and the ``surface-redshift mass", estimated from the light redshift $z_*$ (emitted from the surface to infinity) as
\begin{equation}
M_{\rm sr} \equiv \frac{r_*}{2 G_N} \frac{z_* (2+z_*)}{(1+z_*)^2}.
\end{equation}
Although all of them coincide in GR (except, obviously for the proper mass, whose difference to the others yields the binding energy), they are not the same in $f(R)$, due to the simple fact that the scalar degree of freedom does leaks out from the star. One should then be careful to properly define the star surface: does it happen when the baryonic pressure vanishes or when the total fluid pressure (including then the ``curvature fluid") does? 

\subsubsection{Local gravity constraints}

No review (however short) on $f(R)$ would be complete without mentioning (however briefly) the chameleon effect \cite{2004PhRvD..69d4026K}. This mechanism explains how the extra degree of freedom does not propagate too far --- which would have shown up as an extra force in current experiments. 

In a nut shell, the authors show that the effective mass of the scalar field depend on the energy density of the standard matter, due to the aforementioned coupling between them in the Einstein Frame. The larger the average local density, the heavier the field is and the shorter is its range. Since we are surrounded by a dark-matter halo, all the classical solar-system tests will still hold.

\subsection{Summary}

We finish this section by stating how long we have gone so far: one must come up with an ingenious function $f(R)$ --- that could either be completely {\it ad hoc} (aiming then to pick important characteristics and their outcome to observable quantities) or from first principles (e.g, some low-energy limit of a ``quantum gravity" theory). Such function must follow theoretical (self-consistency, non-divergences) and observational constraints (background and perturbed cosmology, local gravity and solar system). A very useful tool is the conformal transformation to the Einstein frame, where the Lagrangian is similar to GR and, therefore, more palatable. 

In the next Section we will open a large parenthesis to review some basic Thermodynamics, but in a not-so-ordinary approach, given by Catastrophe Theory. The final Section will then make the link between the first two. 

\section{Thermodynamics}

In the following subsections we will use the van der Waals gas as a typical description of a system that goes through a first-order phase transition. Iin spite of its analytical simplicity, it bears the fundamental characteristics of such transformation and, at the same time, allows for full analytical calculations. 

\subsection{Phase transitions}

\begin{figure}[t]
\includegraphics{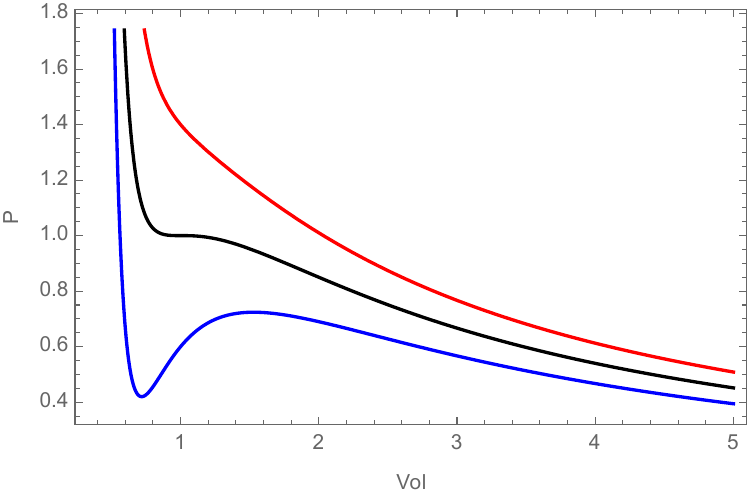}
\caption{Plot of the pressure $\times$ volume for 3 temperatures; the lower (upper) curve corresponds to temperatures below (above) the critical temperature $T_c$. In the middle curve ($T=T_c$), the two extrema (where $dP/dV=0$) coalesce at the inflection point (where $d^2P/dV^2=0$).}
\label{vdwplot}
\end{figure}

The equation of state for a non-ideal gas was introduced by van der Waals in 1873 \cite{vdw}:
\begin{equation}
\left( P + \frac{a}{V^2} \right) \left( V - b\right) = R \, T
\label{vdweq}
\end{equation}
where the constants $a$ and $b$ are introduced to take into account, respectively, a residual interaction between the gas molecules (therefore, decreasing the effective pressure measured) and the volume taken by them. Three of such curves, for different temperatures, are plotted in Fig.~\ref{vdwplot}. One can see that it resembles the ideal gas behavior for large temperatures $T$, as expected. For low temperatures, though, there are two distinguishable regimes: for larger volumes, the pressure is low, while the opposite happens for small volumes. The former region describes a gas-like behavior, while the latter, where the pressure increases fast with a slight decrease in volume, resembles a liquid state, where the interactions between the particles can no longer be neglected.

We recall the reader that, as any equation of state in Thermodynamics, it describes equilibrium configurations, but not necessarily stable ones. For instance, for temperatures  below a critical value $T_c$, there is a volume range where
\begin{equation}
\frac{dP}{dV}>0.
\label{unstable}
\end{equation}
That indicates an unstable equilibrium, since, is the gas was compressed, i.e, its volume would be decreased by an external force, then its pressure would also decrease. In other words, the gas would not try to restore the previous configuration and ultimately it would collapse\footnote{Of course, the opposite argument applies for an expanding gas in this volume range.}. 

A very important piece of information is obtained when one writes the Gibbs Energy $G(P,T)$, whose variation is given by
\begin{equation}
dG(P,T) = V \cdot  dP - S \cdot dT.
\label{dG}
\end{equation}
Therefore, 
\begin{align}
\frac{dG}{dP} &= V \qquad {\rm and}\\
\frac{d^2G}{dP^2} &= \frac{dV}{dP}.
\end{align}
This means that $d^2G/dP^2>0$ in the unstable region. 

One can always write Eq.~(\ref{vdweq}) as
\begin{equation}
V^3 - \left(b + \frac{RT}{P}\right)V^2 - \left(\frac{a}{P}\right) V - \frac{ab}{P}=0.
\label{vdweq2}
\end{equation}
As a cubic equation with real coefficients, there is always (i.e, for any $T$) at least one real root, i.e,  one pressure $P$ and one temperature $T$ for a given volume $V$. On the other hand, for $T<T_c\equiv 8a/(27 b R)$, Eq.~(\ref{vdweq2}) yields three real roots: three values for $V$ (one of them in the aforementioned  unstable region and, therefore, non physical) for a a given pressure $P$ and temperature $T<T_c$. At the critical temperature $Tc$, the system has critical pressure $P_c\equiv a/(27 b^2)$ and critical Volume $V_c\equiv 3 b$. Above $T_c$, there is no difference between the states $A$ and $B$, i.e, one can go from the former to the latter {\it without} going through first-order phase transition.

A simple variable change is usually made so that we can deal only with dimensionless quantities:
\begin{align}
p &\equiv P/P_c - 1 \\
t &\equiv T/T_c -1 \\
x &\equiv V_c/V - 1 ,
\end{align}
upon which Eq.~(\ref{vdweq2}) is written as
\begin{align}
x^3 &+ \alpha x + \beta = 0 \qquad {\rm where}\\
\alpha &\equiv \frac{8 t + p}{3}\\
\beta &\equiv \frac{8 t - 2p}{3}.
\end{align}
The plot of $x$, i.e, the equilibrium configurations as a function of the parameters $\{\alpha,\beta\}$ can be seen in Fig.~\ref{fold}. The region with multiple (to be more precise, three) solutions can be easily seen and its boundary will be determined later on. For now, it suffices to remember that one of the solutions is the unstable one. Outside that region, there is one and only one equilibrium solution.

\subsection{Catastrophe theory}

The equilibrium solutions can be seen as the ones that extremize a suitable potential energy\footnote{Not to be confused with the volume $V$ of the vdW gas.} $V(x)$. A simple  and general expression\footnote{One can get rid of a possible cubic term by a suitable shift on $x$.} that does the job is
\begin{equation}
V(x; \alpha,\beta) \equiv 
\frac{1}{4} x^4 + \frac{\alpha}{2} x^2 + \beta x ,
\label{Vx}
\end{equation}
where we have made explicit the dependence on the control parameters $\{\alpha,\beta\}$. 
In the region with three extrema, there are three Real solutions to Eq.~(\ref{vdweq2}). It is easy to see that one of them is necessarily a local maximum, another one is a local minimum and the last one is the global minimum --- of course, the two minima could be degenerated, but we will come back to this point in a second. Outside that region, there is only one minimum.
The boundary is defined, then, as the location in the parameter space where the local maximum coalesces to the local minimum. Mathematically, it is defined by the so-called {\it Singularity Set}:
\begin{equation}
\frac{d^2V}{dx^2}\bigg|_{x_*} = 0
\label{singset}
\end{equation}
where $x_*$ is one of the extrema, given, of course, by 
\begin{equation}
\frac{dV}{dx}\bigg|_{x_*} = 0,
\label{eqset}
\end{equation}
which is the so-called {\it Equilibrium Set}. The solution of Eqs.~(\ref{singset}) and (\ref{eqset}) is the so-called {\it Bifurcation Set}
\begin{equation}
4 \alpha^3 + 27 \beta^2 =0.
\label{fold}
\end{equation}
In the parameter space $\{\alpha,\beta\}$, the function (\ref{Vx}) has only one extremum (a minimum) outside the curve (\ref{fold}). Inside, there are 3 extrema, 2 of which are minima and 1, a maximum.  On the curve itself, two of them coalesce --- which ones is set by the sign of $\beta$.

\begin{figure}
\centering
\includegraphics[width=0.7\textwidth]{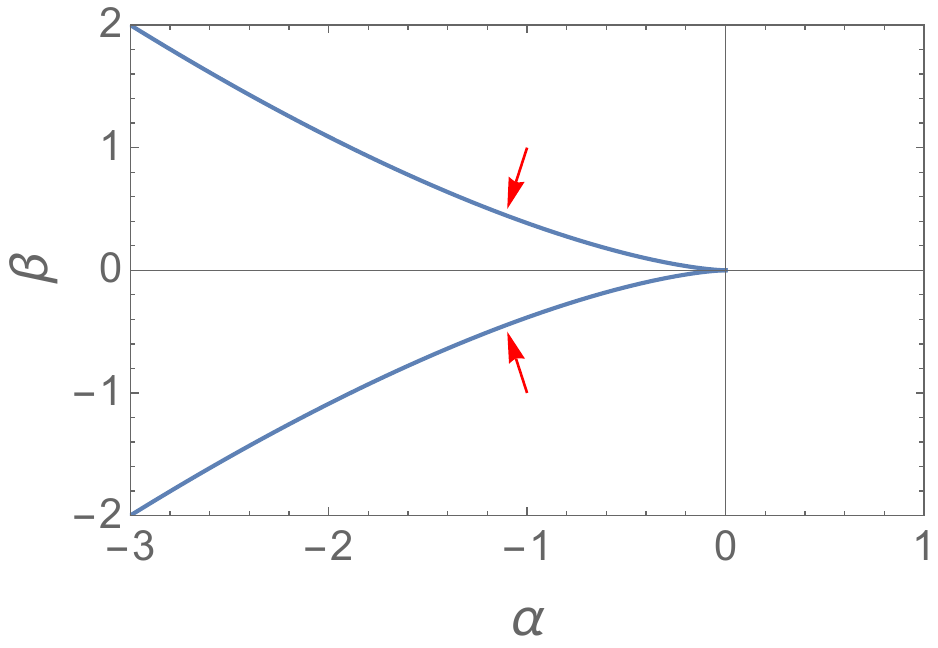}
\begin{picture}(0,0)
\put(-80,150){\includegraphics[width=0.15\textwidth]{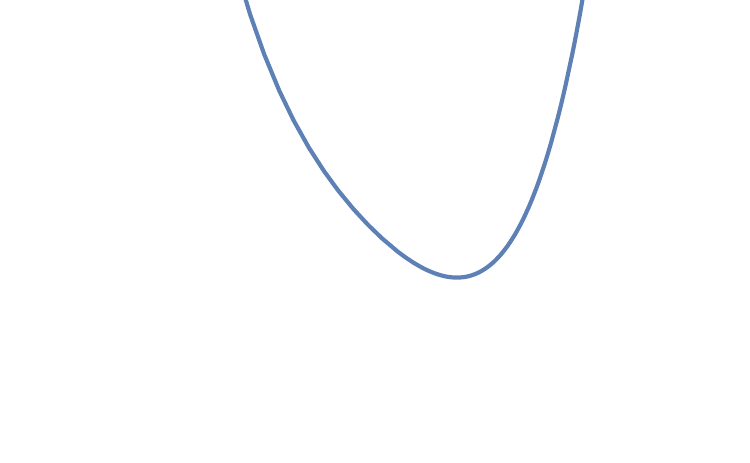}}
\put(-80,50){\includegraphics[width=0.15\textwidth]{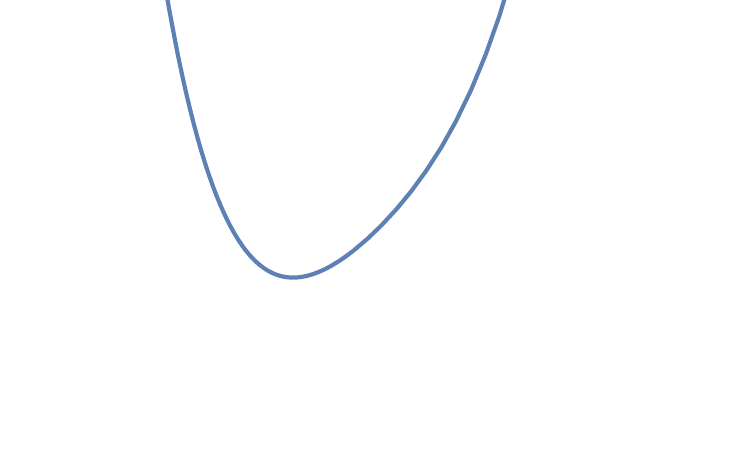}}
\put(-280,130){\includegraphics[width=0.12\textwidth]{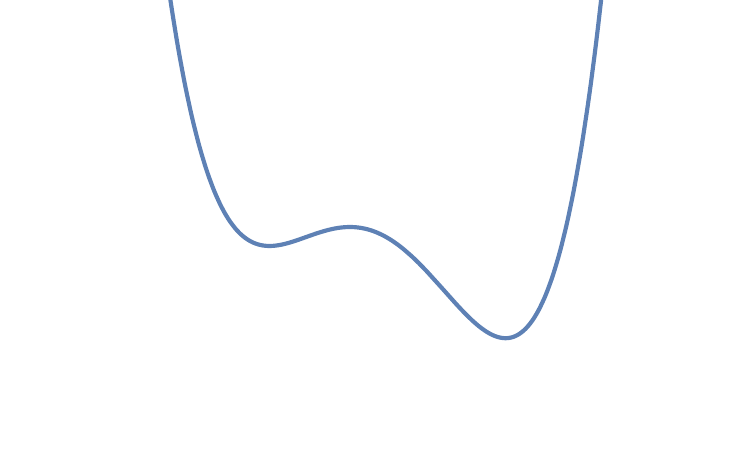}}
\put(-280,80){\includegraphics[width=0.12\textwidth]{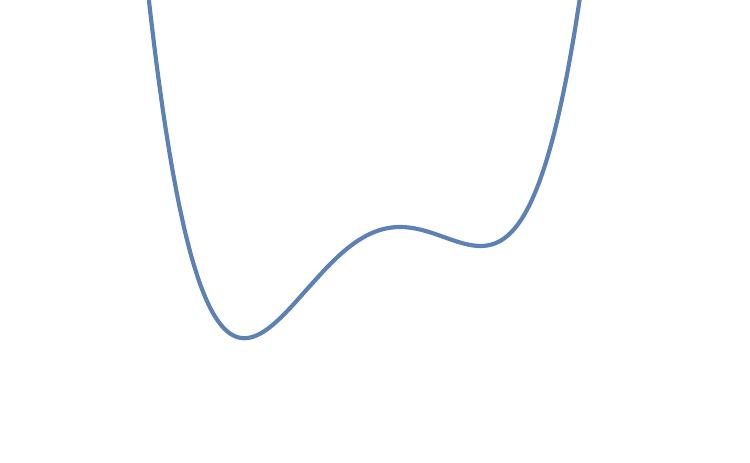}}
\put(-160,170){\includegraphics[width=0.1\textwidth]{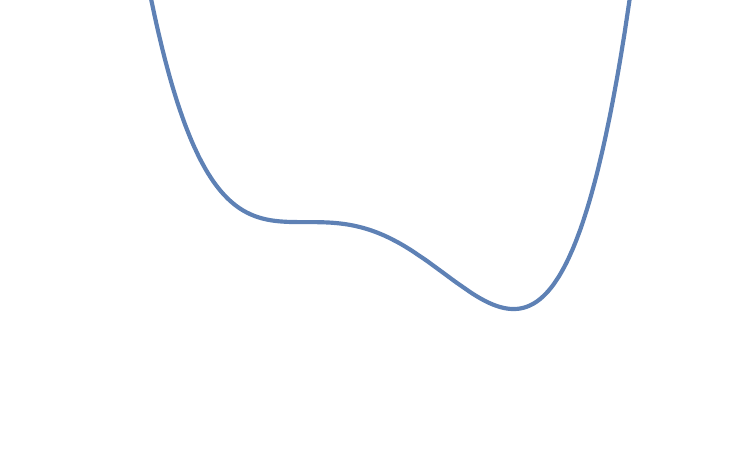}}
\put(-150,50){\includegraphics[width=0.1\textwidth]{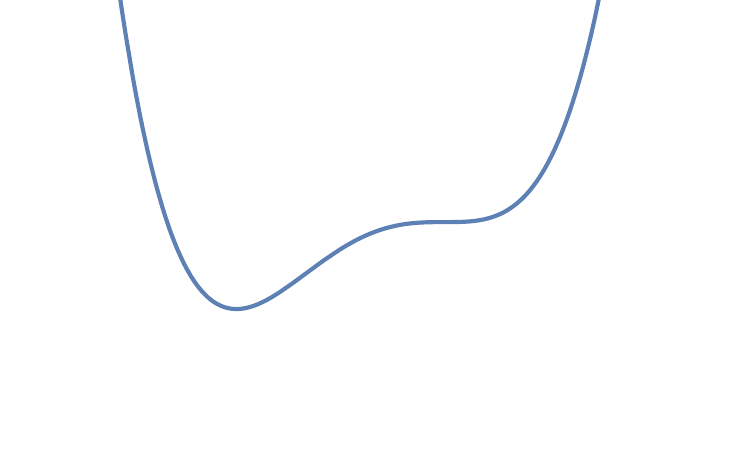}}
\end{picture}
\caption{Plot of the fold given by Eq.~(\ref{fold}) and the corresponding form of the potential energy (\ref{Vx}) in each region. The middle column of insets indicates the form of the potential {\it on} the bifurcation set, as indicated by the arrows. Inside the set, the potential has 3 extrema; outside, only one.}
\label{foldplot}
\end{figure}

\begin{figure}[t]
\includegraphics[width=0.7\textwidth]{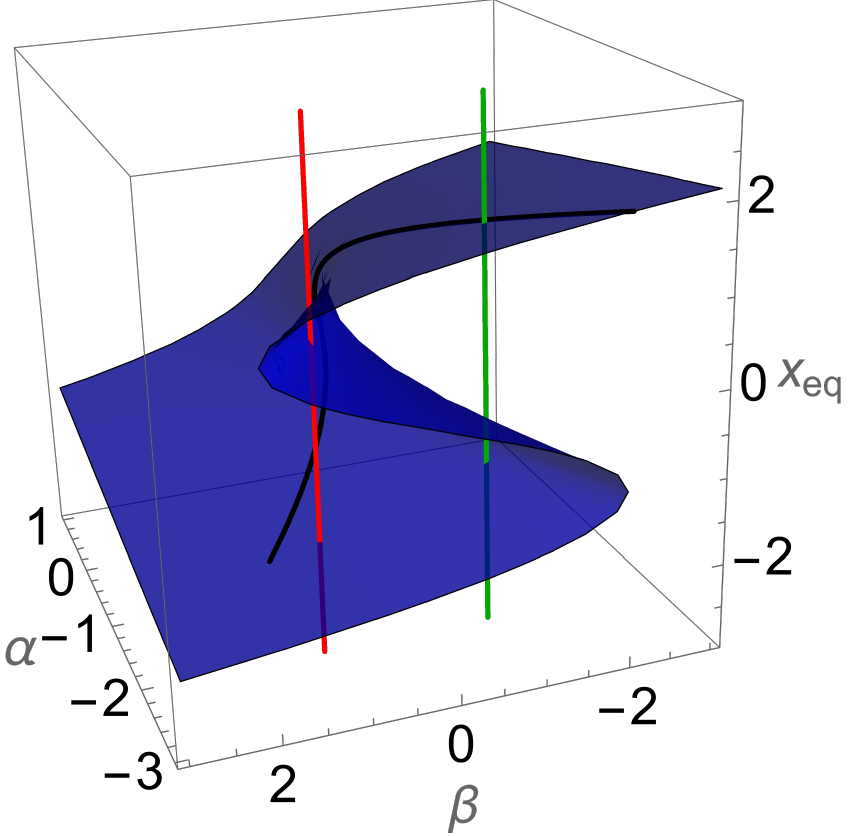}
\caption{Plot of the equilibrium positions, given by Eq.~(\ref{eqset}) as a function of the control parameters $\{\alpha,\beta\}$. The black solid curve is the projection of the fold (\ref{fold}) onto that surface. The vertical red and green straight lines are arbitrary, except for both being inside the fold. Each one indicate the number (3) of equilibrium solutions for each pair $\{\alpha,\beta\}$ and which ones coalesce and which one survives when the fold is crossed.}
\end{figure}

\begin{figure}[t]
\includegraphics[width=0.7\textwidth]{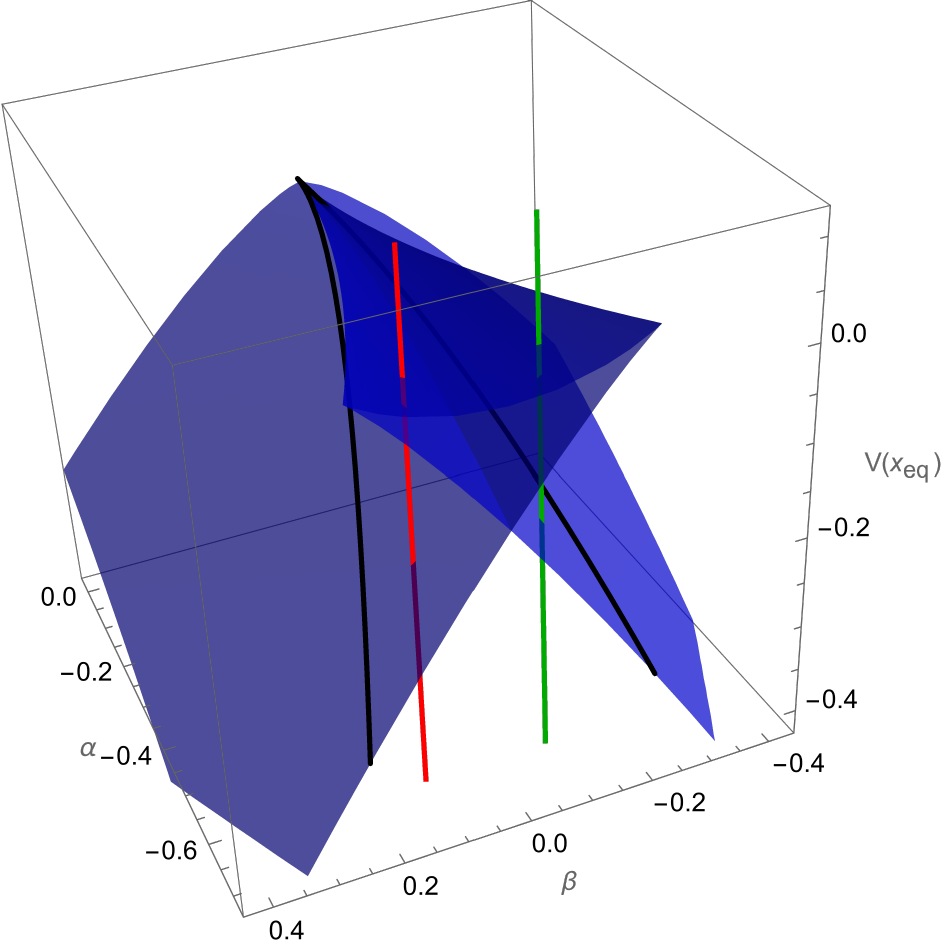}
\caption{Plot of the potential (\ref{Vx}),  calculated at the equilibrium positions $x_{eq}$ given by Eq.~(\ref{eqset}), as a function of the control parameters $\{\alpha,\beta\}$. The black solid curve is the projection of the fold (\ref{fold}) onto that surface. Vertical red and green lines as in the previous figure.}
\label{swallow3D_lines}
\end{figure}

In Fig.~\ref{swallow3D_lines}, one can see the equilibrium surface (i.e, the values of the equilibrium points) as a function of the control parameters $\{\alpha,\beta\}$. The fold is projected onto this surface. One can see that, inside the fold, vertical lines (fixed values of the control parameters) cross the surface three times, each for a given equilibrium solution. If we shift the red (green) line towards larger (smaller) values of $\beta$, i.e., toward the closest fold branch, then the largest (smallest) values of $x_{eq}$ coalesce.  

Accordingly, we can plot the so-called swallowtail curve also as a function of $\{\alpha,\beta\}$ --- see Fig.~\ref{swallow3D_lines}. one can see that the highest $V(x_{eq})$ is always unstable, as expected from a general principle of minimizing the energy. Here, as in the previous plot, when we shift the red (green) vertical line towards the closet fold branch, it is always the equilibrium point corresponding to the highest $V(x_{eq})$ that coalesces with the one corresponding to the intermediate value $V(x_{eq})$, i.e, the local maximum. Outside the fold, only the global minimum survives.

\section{$f(R)$ Theories and Thermodynamics}

Here we follow Ref.~\cite{Peralta_2020}.
We have described the traditional discussion: one comes up with an ingenious non-linear function $f(R)$ (in the Jordan frame) and, via a conformal transformation given by $\phi\equiv f'(R)$, arrives at the Einstein frame, where the physics is described by GR and an extra scalar field $\tilde\phi$ subject to a potential $V(\tilde\phi$), completely defined by the function $f(R)$ chosen at the beginning.

Here, we take the opposite direction: we start at the Einstein frame with the simplest potential and look for the corresponding $f(R)$ in the Jordan frame. We assume
\begin{equation}
V(\tilde\phi) = \frac{1}{2} \tilde m_\phi^2 (\tilde\phi -a)^2 + \Lambda.
\label{singlewell}
\end{equation}
For now, the extra (constant) parameters $a$ and $\Lambda$ are simply a matter of generalizing the potential $V$, while keeping it as simple as possible. One might argue against the introduction of a cosmological constant $\Lambda$, since it seems to be incompatible with one of the core motivations for modifying GR. It will be, however, essential to what follows. 

\begin{figure}
\center
\includegraphics[width=0.45\textwidth]{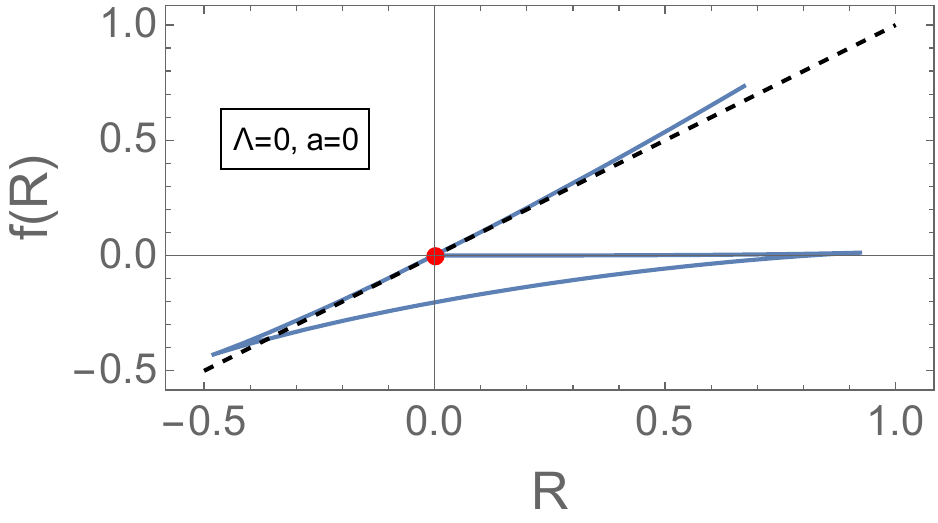}
\includegraphics[width=0.45\textwidth]{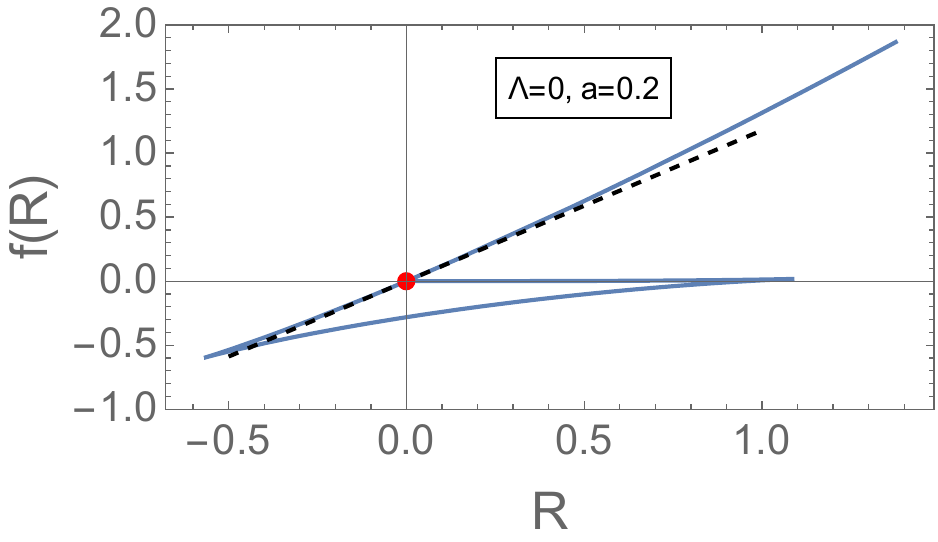}
\includegraphics[width=0.45\textwidth]{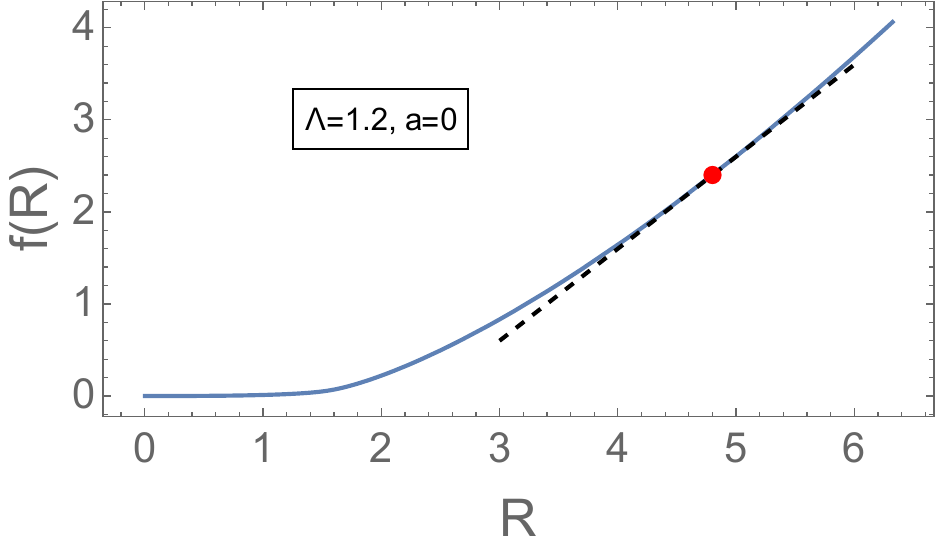}
\includegraphics[width=0.45\textwidth]{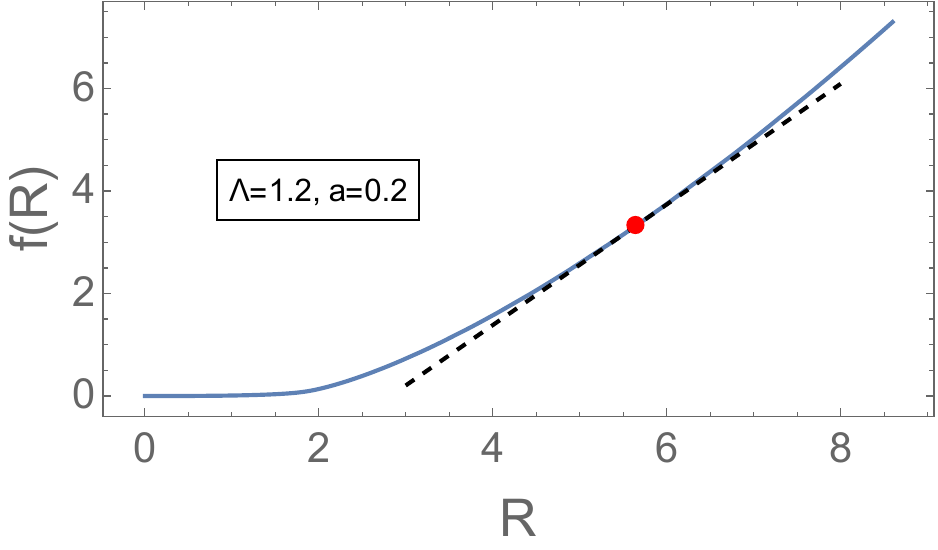}
\caption{Plots of $f(R) \times R$ for different values of the free parameters $\Lambda$ and $a$, as indicated in each panel. Note the absence of the unstable region (where $f''<0$) for large $\Lambda$. Different values of $a$ simply change the axis scales.}
\label{swallow}
\end{figure}

Following a previously established procedure \cite{1994PhRvD..50.5039M}, one arrives at the following parametric expressions, where $\beta\equiv \sqrt{2/3}$:
\begin{align}
\label{fphi}
f(\tilde\phi) &= {\rm e}^{2 \beta \tilde\phi} \left[2V(\tilde\phi) + 2 \beta^{-1}  \frac{{\rm d} V(\tilde\phi)}{d\tilde\phi} \right] \quad {\rm and}\\
\label{Rphi}
R(\tilde\phi) &= \, {\rm e}^{\beta \tilde\phi} \,\left[4 V(\tilde\phi) + 2 \beta^{-1} \frac{{\rm d} V(\tilde\phi)}{d\tilde\phi} \right].
\end{align}
We have plotted the above expression for the potential (\ref{singlewell}) in Fig.~\ref{swallow}, for different values of $\{\Lambda,a\}$.

We first notice a few important characteristics: First and more obvious: the function $f(R)$ is multi valued. Its actual value depends on $\tilde\phi$ while it evolves towards the bottom of its potential. Secondly, $f(R)>0$ for all $R$ and $f''<0$ only in the middle, lower branch, indicating instability of the theory. Thirdly, that very branch is absent for $\Lambda\geqslant \Lambda_c = 15/16$, which indicates it is a control parameter, such as temperature. 

\begin{figure}[t]
\center
\includegraphics[width=0.42\textwidth]{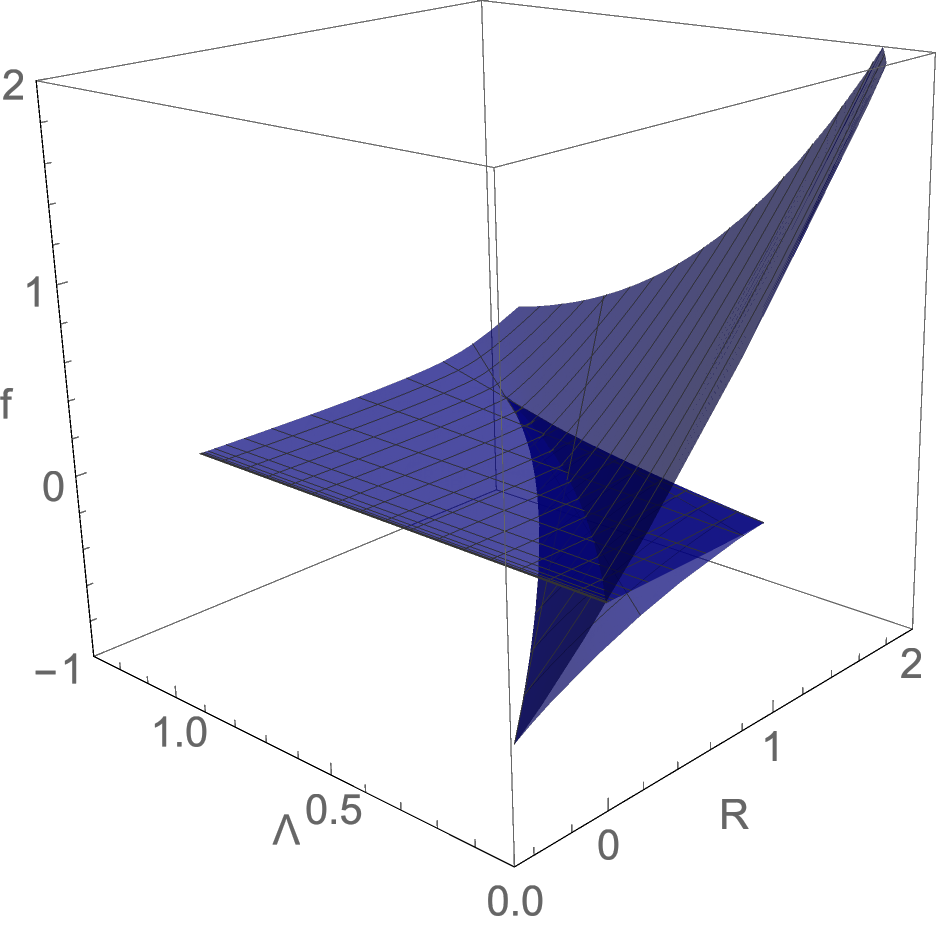}
\includegraphics[width=0.42\textwidth]{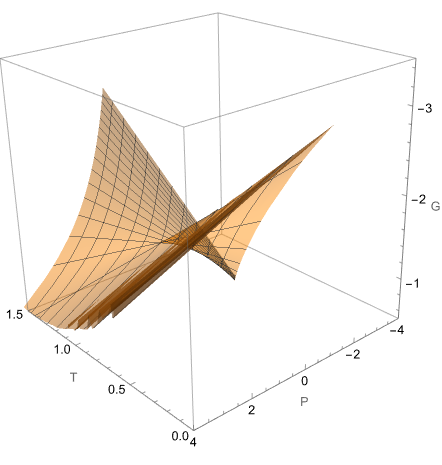}
\caption{Plots of {\bf (left panel)} $f(R,\Lambda)$, given by Eqs.~(\ref{fphi}) and (\ref{Rphi}) with $a=\Lambda=0$, and {\bf (right panel)} 
 $G(P,T)$ for de vdW gas. Two of the axes in the latter are inverted to allow a clear comparison to the former.}
\label{3Dswallow}
\end{figure}

Besides, if we look at Eq.~(\ref{Lf}) as the Lagrangian of a relativistic fluid, described by its pressure $P$, the final step is almost automatic:  to make the correspondence between $R$ itself and the Gibbs Energy $G$. This is clearly seen if we plot the 3D behavior of $f(R,\Lambda)$ and the Gibbs Energy $G(P,T)$) in Fig.~\ref{3Dswallow}.

The effective volume $V$ is the variable ``canonically conjugated" to the effective pressure $P$, i.e, since
\begin{equation}
dG(P,T) = V \cdot  dP - S \cdot dT.
\label{dG2}
\end{equation}
We have, therefore, the full expressions for the Gibbs Energy $G$, the pressure $P$ and the Volume $V$, in terms of the field $\tilde\phi$:
\begin{align}
G &= e^{\beta  \tilde\phi } \left(\frac{2 (\tilde\phi -a) [\beta  (\tilde\phi -a)+1]}{\beta }+4 T\right)
\label{Gibbs}
\\
P &= e^{2 \beta  \tilde\phi } \left(\frac{2 (\tilde\phi -a)}{\beta }+(\tilde\phi -a)^2+2 T\right) 
\label{Pphi}\\
V &= \exp(-\beta \tilde\phi) \quad \Leftrightarrow \quad \tilde\phi = -\frac{1}{\beta}\log(V),
\label{Vphi}
\end{align}
which allows us to plot the 3D behavior of $G(P,T)$ in Fig.~\ref{3Dswallow}. The unstable region, where $d^2G/dP^2>0$ (notice that the $G$ axis points down) corresponds to higher values of the Gibbs Energy for a fixed $T$.

One can also calculate the Helmholtz energy 
\begin{align}
F(T,V) &\equiv G - P \cdot V  \\
& = \frac{1}{V}\bigg[\left(a+\frac{1}{\beta }\log V\right)^2+2 T\bigg]
\end{align}
and the Entropy 
\begin{align}
S (T,V) &\equiv - \left.\frac{\partial F}{\partial T}\right|_V  \\
& = -\frac{2}{V}.
\end{align}
There are two important notes from the result above. First, we recall the reader that, as the field $\tilde\phi$ sets itself at the bottom of its potential ($\tilde\phi= a$), the volume assumes its minimum value $V\to \exp(-\beta a)$), which corresponds to a decrease in Entropy. That indicates that the system is not complete, i.e, we are not looking only at the full system. Indeed, we have not included matter nor radiation, which should then absorb the latent heat released in the phase transition from inflation to a radiation-dominated universe (see discussion further down). Secondly, we must correct its negative value, since it prevents a physical interpretation in terms of the number of accessible states. For that, it suffices to add an extra term in Eq.~(\ref{Gibbs}) and redefine the Gibbs Energy it as
\begin{equation}
G = e^{\beta  \tilde\phi } \left(\frac{2 (\tilde\phi -a) [\beta  (\tilde\phi -a)+1]}{\beta }+4 T\right)- 2 T e^{\beta a},
\label{Gibbs2}
\end{equation}
which does not spoil the previous results.

\begin{figure}[t]
\center
\includegraphics[width=0.7\textwidth]{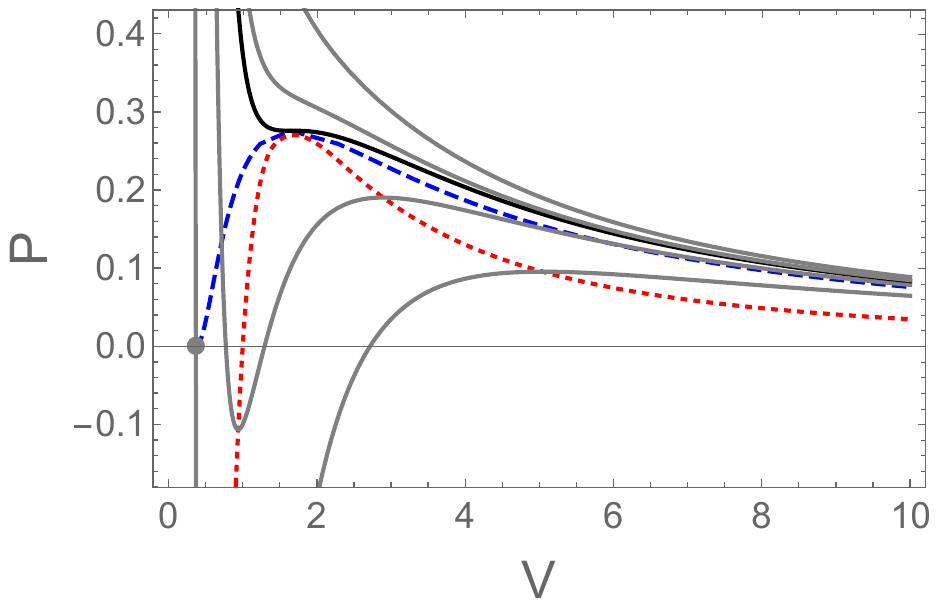}
\caption{Plot of the effective pressure $P$ as a function of the effective volume $V$, for $a=a_*\equiv 1/\beta$ and different values of temperature: $T=T_c\equiv 15/16$ (solid thick black); lower (higher) curves, in solid thin gray, correspond to lower (higher) temperatures. The lowest one (bottom not shown) corresponds to $\Lambda\equiv T = 0$. The {\bf spinodal curve} is plotted in dotted red. The {\bf binodal curve} is plotted in dashed blue. The gray circle indicates the final configuration ($\phi=a$) for the $T=0$ case (higher temperatures correspond to  higher final pressures).}
\label{sbin}
\end{figure}

From Eqs.~(\ref{Pphi}) and (\ref{Vphi}), we can write the equation of state, i.e, the relation between the pressure, volume and temperature:
\begin{equation}
P(V,T) = \frac{\beta  \left(a^2 \beta -2 a+2 \beta  T\right)+(2 a \beta -2 +\log V )\log V }{\beta ^2 V^2}.
\label{PV}
\end{equation}
The behavior of $P(V,T)$ for five different values of $T$ is shown in Fig.~\ref{sbin}, which bears strong resemblance to a vdW gas\footnote{Nevertheless, here one obtains $P\propto T V^{-2}$ in the high-temperature limit, instead of the standard ideal-gas behavior $P \propto T V^{-1}$.}. Even though the equations of state are not exactly the same, they do describe the same phenomena, as we will now see. 

The interesting Physics happens around the region where $dP/dV>0$, i.e, the unstable one. Fig.~\ref{sbin} shows also the binodal and spinodal lines. The former indicates the limit of existence of unstable configurations (which are exactly the Maxwell Construction, also known as the equal-area principle), while below the latter there are unstable states. 

One can map the initial conditions that satisfy the slow-roll conditions (large negative $\tilde\phi$) into the right-most region in Fig.~\ref{sbin}. If we follow the $(T=0)$-curve, the system starts below the binodal line, which is exactly what is needed: the inflationary era should be metastable --- i.e, it should last a few ($\sim 60$) e-folds, not too short (as it would be for an unstable state) and not too long (as in a stable state). The final state, when $\tilde\phi$ is oscillating around the bottom of its potential, corresponds to the opposite end of the plot (small volume), in a different phase --- for the case at hand, with a quadratic potential (\ref{singlewell}), the average value of the equation-of-state parameter for the field $\tilde\phi$ is $\langle \tilde\omega_\phi \rangle \equiv \langle \tilde p_\phi/\tilde\rho_\phi \rangle \approx 0$, i.e, a matter-dominated universe. 
We refer the reader to Ref.~\cite{Peralta_2020} to a numerical description and the full Thermodynamic correspondence.

\section{Final summary}

We have seen a strong correspondence between $f(R)$ theories (in the metric approach) and a first-order phase transition, as described by Catastrophe Theory. That analogy has only been investigated in the inflationary scenario, so far. 

Current work in under way for the (p)reheating mechanism that leads to an homogeneous hot dense universe compatible to a radiation-dominated phase followed by a matter-dominated one, that GR describes so nicely. Of course, relativistic stars and perturbation growth are also part of the next steps, as well as more complex potentials in the Einstein frame.

I would like to thank Mario Novello for keeping such a successful School (along with its international counterpart, the Brazilian School of Cosmology and Gravitation) alive and thriving for so many years. I acknowledge financial support from FAPERJ.



\begin{thebibliography}{26}%
\makeatletter
\providecommand \@ifxundefined [1]{%
 \@ifx{#1\undefined}
}%
\providecommand \@ifnum [1]{%
 \ifnum #1\expandafter \@firstoftwo
 \else \expandafter \@secondoftwo
 \fi
}%
\providecommand \@ifx [1]{%
 \ifx #1\expandafter \@firstoftwo
 \else \expandafter \@secondoftwo
 \fi
}%
\providecommand \natexlab [1]{#1}%
\providecommand \enquote  [1]{``#1''}%
\providecommand \bibnamefont  [1]{#1}%
\providecommand \bibfnamefont [1]{#1}%
\providecommand \citenamefont [1]{#1}%
\providecommand \href@noop [0]{\@secondoftwo}%
\providecommand \href [0]{\begingroup \@sanitize@url \@href}%
\providecommand \@href[1]{\@@startlink{#1}\@@href}%
\providecommand \@@href[1]{\endgroup#1\@@endlink}%
\providecommand \@sanitize@url [0]{\catcode `\\12\catcode `\$12\catcode
  `\&12\catcode `\#12\catcode `\^12\catcode `\_12\catcode `\%12\relax}%
\providecommand \@@startlink[1]{}%
\providecommand \@@endlink[0]{}%
\providecommand \url  [0]{\begingroup\@sanitize@url \@url }%
\providecommand \@url [1]{\endgroup\@href {#1}{\urlprefix }}%
\providecommand \urlprefix  [0]{URL }%
\providecommand \Eprint [0]{\href }%
\providecommand \doibase [0]{http://dx.doi.org/}%
\providecommand \selectlanguage [0]{\@gobble}%
\providecommand \bibinfo  [0]{\@secondoftwo}%
\providecommand \bibfield  [0]{\@secondoftwo}%
\providecommand \translation [1]{[#1]}%
\providecommand \BibitemOpen [0]{}%
\providecommand \bibitemStop [0]{}%
\providecommand \bibitemNoStop [0]{.\EOS\space}%
\providecommand \EOS [0]{\spacefactor3000\relax}%
\providecommand \BibitemShut  [1]{\csname bibitem#1\endcsname}%
\let\auto@bib@innerbib\@empty
\bibitem [{\citenamefont {{Sotiriou}}\ and\ \citenamefont
  {{Faraoni}}(2010)}]{Sotiriou_2010}%
  \BibitemOpen
  \bibfield  {author} {\bibinfo {author} {\bibfnamefont {T.~P.}\ \bibnamefont
  {{Sotiriou}}}\ and\ \bibinfo {author} {\bibfnamefont {V.}~\bibnamefont
  {{Faraoni}}},\ }\href {\doibase 10.1103/RevModPhys.82.451} {\bibfield
  {journal} {\bibinfo  {journal} {Reviews of Modern Physics}\ }\textbf
  {\bibinfo {volume} {82}},\ \bibinfo {pages} {451} (\bibinfo {year} {2010})},\
  \Eprint {http://arxiv.org/abs/0805.1726} {arXiv:0805.1726 [gr-qc]}
  \BibitemShut {NoStop}%
\bibitem [{\citenamefont {Sotiriou}(2007)}]{sotiriou2007modified}%
  \BibitemOpen
  \bibfield  {author} {\bibinfo {author} {\bibfnamefont {T.~P.}\ \bibnamefont
  {Sotiriou}},\ }\href@noop {} {\enquote {\bibinfo {title} {Modified actions
  for gravity: Theory and phenomenology},}\ } (\bibinfo {year} {2007}),\
  \Eprint {http://arxiv.org/abs/0710.4438} {arXiv:0710.4438 [gr-qc]}
  \BibitemShut {NoStop}%
\bibitem [{\citenamefont {{De Felice}}\ and\ \citenamefont
  {{Tsujikawa}}(2010)}]{2010LRR....13....3D}%
  \BibitemOpen
  \bibfield  {author} {\bibinfo {author} {\bibfnamefont {A.}~\bibnamefont {{De
  Felice}}}\ and\ \bibinfo {author} {\bibfnamefont {S.}~\bibnamefont
  {{Tsujikawa}}},\ }\href {\doibase 10.12942/lrr-2010-3} {\bibfield  {journal}
  {\bibinfo  {journal} {Living Reviews in Relativity}\ }\textbf {\bibinfo
  {volume} {13}},\ \bibinfo {eid} {3} (\bibinfo {year} {2010})},\ \Eprint
  {http://arxiv.org/abs/1002.4928} {arXiv:1002.4928 [gr-qc]} \BibitemShut
  {NoStop}%
\bibitem [{\citenamefont {{Amendola}}\ and\ \citenamefont
  {{Tsujikawa}}(2015)}]{2015daen.book.....A}%
  \BibitemOpen
  \bibfield  {author} {\bibinfo {author} {\bibfnamefont {L.}~\bibnamefont
  {{Amendola}}}\ and\ \bibinfo {author} {\bibfnamefont {S.}~\bibnamefont
  {{Tsujikawa}}},\ }\href@noop {} {\emph {\bibinfo {title} {{Dark Energy}}}}\
  (\bibinfo {year} {2015})\BibitemShut {NoStop}%
\bibitem [{\citenamefont {{Callen}}(1985)}]{1985tait.book.....C}%
  \BibitemOpen
  \bibfield  {author} {\bibinfo {author} {\bibfnamefont {H.~B.}\ \bibnamefont
  {{Callen}}},\ }\href@noop {} {\emph {\bibinfo {title} {{Thermodynamics and an
  Introduction to Thermostatistics, 2nd Edition}}}}\ (\bibinfo {year}
  {1985})\BibitemShut {NoStop}%
\bibitem [{\citenamefont {Gilmore}(1981)}]{gilmore-1981a}%
  \BibitemOpen
  \bibfield  {author} {\bibinfo {author} {\bibfnamefont {R.}~\bibnamefont
  {Gilmore}},\ }\href@noop {} {\emph {\bibinfo {title} {Catastrophe Theory for
  Scientists and Engineers}}}\ (\bibinfo  {publisher} {Dover},\ \bibinfo
  {address} {New York},\ \bibinfo {year} {1981})\BibitemShut {NoStop}%
\bibitem [{\citenamefont {Poston}\ and\ \citenamefont
  {Stewart}(1996)}]{poston1996catastrophe}%
  \BibitemOpen
  \bibfield  {author} {\bibinfo {author} {\bibfnamefont {T.}~\bibnamefont
  {Poston}}\ and\ \bibinfo {author} {\bibfnamefont {I.}~\bibnamefont
  {Stewart}},\ }\href {https://books.google.com.br/books?id=7Zm5zTh8rLAC}
  {\emph {\bibinfo {title} {Catastrophe Theory and Its Applications}}},\ Dover
  books on mathematics\ (\bibinfo  {publisher} {Dover Publications},\ \bibinfo
  {year} {1996})\BibitemShut {NoStop}%
\bibitem [{\citenamefont {Saunders}(1980)}]{saunders_1980}%
  \BibitemOpen
  \bibfield  {author} {\bibinfo {author} {\bibfnamefont {P.~T.}\ \bibnamefont
  {Saunders}},\ }\href {\doibase 10.1017/CBO9781139171533} {\emph {\bibinfo
  {title} {An Introduction to Catastrophe Theory}}}\ (\bibinfo  {publisher}
  {Cambridge University Press},\ \bibinfo {year} {1980})\BibitemShut {NoStop}%
\bibitem [{\citenamefont {{Milgrom}}(1983)}]{1983ApJ...270..371M}%
  \BibitemOpen
  \bibfield  {author} {\bibinfo {author} {\bibfnamefont {M.}~\bibnamefont
  {{Milgrom}}},\ }\href {\doibase 10.1086/161131} {\bibfield  {journal}
  {\bibinfo  {journal} {\apj}\ }\textbf {\bibinfo {volume} {270}},\ \bibinfo
  {pages} {371} (\bibinfo {year} {1983})}\BibitemShut {NoStop}%
\bibitem [{\citenamefont {Marra}\ \emph {et~al.}(2020)\citenamefont {Marra},
  \citenamefont {Rodrigues},\ and\ \citenamefont {de Almeida}}]{Marra_2020}%
  \BibitemOpen
  \bibfield  {author} {\bibinfo {author} {\bibfnamefont {V.}~\bibnamefont
  {Marra}}, \bibinfo {author} {\bibfnamefont {D.~C.}\ \bibnamefont
  {Rodrigues}}, \ and\ \bibinfo {author} {\bibfnamefont {A.~O.~F.}\
  \bibnamefont {de Almeida}},\ }\href {\doibase 10.1093/mnras/staa890}
  {\bibfield  {journal} {\bibinfo  {journal} {Monthly Notices of the Royal
  Astronomical Society}\ }\textbf {\bibinfo {volume} {494}},\ \bibinfo {pages}
  {2875–2885} (\bibinfo {year} {2020})}\BibitemShut {NoStop}%
\bibitem [{\citenamefont {{Olmo}}(2011)}]{2011IJMPD..20..413O}%
  \BibitemOpen
  \bibfield  {author} {\bibinfo {author} {\bibfnamefont {G.~J.}\ \bibnamefont
  {{Olmo}}},\ }\href {\doibase 10.1142/S0218271811018925} {\bibfield  {journal}
  {\bibinfo  {journal} {International Journal of Modern Physics D}\ }\textbf
  {\bibinfo {volume} {20}},\ \bibinfo {pages} {413} (\bibinfo {year} {2011})},\
  \Eprint {http://arxiv.org/abs/1101.3864} {arXiv:1101.3864 [gr-qc]}
  \BibitemShut {NoStop}%
\bibitem [{\citenamefont {{Sotiriou}}\ and\ \citenamefont
  {{Liberati}}(2007)}]{2007AnPhy.322..935S}%
  \BibitemOpen
  \bibfield  {author} {\bibinfo {author} {\bibfnamefont {T.~P.}\ \bibnamefont
  {{Sotiriou}}}\ and\ \bibinfo {author} {\bibfnamefont {S.}~\bibnamefont
  {{Liberati}}},\ }\href {\doibase 10.1016/j.aop.2006.06.002} {\bibfield
  {journal} {\bibinfo  {journal} {Annals of Physics}\ }\textbf {\bibinfo
  {volume} {322}},\ \bibinfo {pages} {935} (\bibinfo {year} {2007})},\ \Eprint
  {http://arxiv.org/abs/gr-qc/0604006} {arXiv:gr-qc/0604006 [gr-qc]}
  \BibitemShut {NoStop}%
\bibitem [{\citenamefont {Hernandez-Arboleda}\ \emph
  {et~al.}(2023)\citenamefont {Hernandez-Arboleda}, \citenamefont {Rodrigues},
  \citenamefont {Toniato},\ and\ \citenamefont
  {Wojnar}}]{hernandezarboleda2023palatini}%
  \BibitemOpen
  \bibfield  {author} {\bibinfo {author} {\bibfnamefont {A.}~\bibnamefont
  {Hernandez-Arboleda}}, \bibinfo {author} {\bibfnamefont {D.~C.}\ \bibnamefont
  {Rodrigues}}, \bibinfo {author} {\bibfnamefont {J.~D.}\ \bibnamefont
  {Toniato}}, \ and\ \bibinfo {author} {\bibfnamefont {A.}~\bibnamefont
  {Wojnar}},\ }\href@noop {} {\enquote {\bibinfo {title} {Palatini $f(r)$
  gravity tests in the weak field limit: Solar system, seismology and
  galaxies},}\ } (\bibinfo {year} {2023}),\ \Eprint
  {http://arxiv.org/abs/2306.04475} {arXiv:2306.04475 [gr-qc]} \BibitemShut
  {NoStop}%
\bibitem [{\citenamefont {{Kaiser}}(1995)}]{1995PhRvD..52.4295K}%
  \BibitemOpen
  \bibfield  {author} {\bibinfo {author} {\bibfnamefont {D.~I.}\ \bibnamefont
  {{Kaiser}}},\ }\href {\doibase 10.1103/PhysRevD.52.4295} {\bibfield
  {journal} {\bibinfo  {journal} {\prd}\ }\textbf {\bibinfo {volume} {52}},\
  \bibinfo {pages} {4295} (\bibinfo {year} {1995})},\ \Eprint
  {http://arxiv.org/abs/astro-ph/9408044} {arXiv:astro-ph/9408044 [astro-ph]}
  \BibitemShut {NoStop}%
\bibitem [{\citenamefont {Pogosian}\ and\ \citenamefont
  {Silvestri}(2008)}]{PhysRevD.77.023503}%
  \BibitemOpen
  \bibfield  {author} {\bibinfo {author} {\bibfnamefont {L.}~\bibnamefont
  {Pogosian}}\ and\ \bibinfo {author} {\bibfnamefont {A.}~\bibnamefont
  {Silvestri}},\ }\href {\doibase 10.1103/PhysRevD.77.023503} {\bibfield
  {journal} {\bibinfo  {journal} {Phys. Rev. D}\ }\textbf {\bibinfo {volume}
  {77}},\ \bibinfo {pages} {023503} (\bibinfo {year} {2008})}\BibitemShut
  {NoStop}%
\bibitem [{\citenamefont {Amendola}\ \emph {et~al.}(2007)\citenamefont
  {Amendola}, \citenamefont {Gannouji}, \citenamefont {Polarski},\ and\
  \citenamefont {Tsujikawa}}]{Amendola_2007}%
  \BibitemOpen
  \bibfield  {author} {\bibinfo {author} {\bibfnamefont {L.}~\bibnamefont
  {Amendola}}, \bibinfo {author} {\bibfnamefont {R.}~\bibnamefont {Gannouji}},
  \bibinfo {author} {\bibfnamefont {D.}~\bibnamefont {Polarski}}, \ and\
  \bibinfo {author} {\bibfnamefont {S.}~\bibnamefont {Tsujikawa}},\ }\href
  {\doibase 10.1103/physrevd.75.083504} {\bibfield  {journal} {\bibinfo
  {journal} {Physical Review D}\ }\textbf {\bibinfo {volume} {75}} (\bibinfo
  {year} {2007}),\ 10.1103/physrevd.75.083504}\BibitemShut {NoStop}%
\bibitem [{\citenamefont {Herrera}\ \emph {et~al.}(2017)\citenamefont
  {Herrera}, \citenamefont {Waga},\ and\ \citenamefont
  {Jor{\'{a}}s}}]{Herrera_2017}%
  \BibitemOpen
  \bibfield  {author} {\bibinfo {author} {\bibfnamefont {D.}~\bibnamefont
  {Herrera}}, \bibinfo {author} {\bibfnamefont {I.}~\bibnamefont {Waga}}, \
  and\ \bibinfo {author} {\bibfnamefont {S.}~\bibnamefont {Jor{\'{a}}s}},\
  }\href {\doibase 10.1103/physrevd.95.064029} {\bibfield  {journal} {\bibinfo
  {journal} {Physical Review D}\ }\textbf {\bibinfo {volume} {95}} (\bibinfo
  {year} {2017}),\ 10.1103/physrevd.95.064029}\BibitemShut {NoStop}%
\bibitem [{\citenamefont {{Bertschinger}}\ and\ \citenamefont
  {{Zukin}}(2008)}]{2008PhRvD..78b4015B}%
  \BibitemOpen
  \bibfield  {author} {\bibinfo {author} {\bibfnamefont {E.}~\bibnamefont
  {{Bertschinger}}}\ and\ \bibinfo {author} {\bibfnamefont {P.}~\bibnamefont
  {{Zukin}}},\ }\href {\doibase 10.1103/PhysRevD.78.024015} {\bibfield
  {journal} {\bibinfo  {journal} {\prd}\ }\textbf {\bibinfo {volume} {78}},\
  \bibinfo {eid} {024015} (\bibinfo {year} {2008})},\ \Eprint
  {http://arxiv.org/abs/0801.2431} {arXiv:0801.2431 [astro-ph]} \BibitemShut
  {NoStop}%
\bibitem [{\citenamefont {{Borisov}}\ \emph {et~al.}(2012)\citenamefont
  {{Borisov}}, \citenamefont {{Jain}},\ and\ \citenamefont
  {{Zhang}}}]{2012PhRvD..85f3518B}%
  \BibitemOpen
  \bibfield  {author} {\bibinfo {author} {\bibfnamefont {A.}~\bibnamefont
  {{Borisov}}}, \bibinfo {author} {\bibfnamefont {B.}~\bibnamefont {{Jain}}}, \
  and\ \bibinfo {author} {\bibfnamefont {P.}~\bibnamefont {{Zhang}}},\ }\href
  {\doibase 10.1103/PhysRevD.85.063518} {\bibfield  {journal} {\bibinfo
  {journal} {\prd}\ }\textbf {\bibinfo {volume} {85}},\ \bibinfo {eid} {063518}
  (\bibinfo {year} {2012})},\ \Eprint {http://arxiv.org/abs/1102.4839}
  {arXiv:1102.4839 [astro-ph.CO]} \BibitemShut {NoStop}%
\bibitem [{\citenamefont {Batista}(2021)}]{Batista_2021}%
  \BibitemOpen
  \bibfield  {author} {\bibinfo {author} {\bibfnamefont {R.~C.}\ \bibnamefont
  {Batista}},\ }\href {\doibase 10.3390/universe8010022} {\bibfield  {journal}
  {\bibinfo  {journal} {Universe}\ }\textbf {\bibinfo {volume} {8}},\ \bibinfo
  {pages} {22} (\bibinfo {year} {2021})}\BibitemShut {NoStop}%
\bibitem [{\citenamefont {{Pretel}}\ \emph {et~al.}(2023)\citenamefont
  {{Pretel}}, \citenamefont {{Jor{\'a}s}}, \citenamefont {{Reis}},
  \citenamefont {{Duarte}},\ and\ \citenamefont
  {{Arba{\~n}il}}}]{2023arXiv230800203P}%
  \BibitemOpen
  \bibfield  {author} {\bibinfo {author} {\bibfnamefont {J.~M.~Z.}\
  \bibnamefont {{Pretel}}}, \bibinfo {author} {\bibfnamefont {S.~E.}\
  \bibnamefont {{Jor{\'a}s}}}, \bibinfo {author} {\bibfnamefont {R.~R.~R.}\
  \bibnamefont {{Reis}}}, \bibinfo {author} {\bibfnamefont {S.~B.}\
  \bibnamefont {{Duarte}}}, \ and\ \bibinfo {author} {\bibfnamefont {J.~D.~V.}\
  \bibnamefont {{Arba{\~n}il}}},\ }\href {\doibase 10.48550/arXiv.2308.00203}
  {\bibfield  {journal} {\bibinfo  {journal} {arXiv e-prints}\ ,\ \bibinfo
  {eid} {arXiv:2308.00203}} (\bibinfo {year} {2023})},\ \Eprint
  {http://arxiv.org/abs/2308.00203} {arXiv:2308.00203 [gr-qc]} \BibitemShut
  {NoStop}%
\bibitem [{\citenamefont {{Sbis{\`a}}}\ \emph {et~al.}(2020)\citenamefont
  {{Sbis{\`a}}}, \citenamefont {{Baqui}}, \citenamefont {{Miranda}},
  \citenamefont {{Jor{\'a}s}},\ and\ \citenamefont
  {{Piattella}}}]{2020PDU....2700411S}%
  \BibitemOpen
  \bibfield  {author} {\bibinfo {author} {\bibfnamefont {F.}~\bibnamefont
  {{Sbis{\`a}}}}, \bibinfo {author} {\bibfnamefont {P.~O.}\ \bibnamefont
  {{Baqui}}}, \bibinfo {author} {\bibfnamefont {T.}~\bibnamefont {{Miranda}}},
  \bibinfo {author} {\bibfnamefont {S.~E.}\ \bibnamefont {{Jor{\'a}s}}}, \ and\
  \bibinfo {author} {\bibfnamefont {O.~F.}\ \bibnamefont {{Piattella}}},\
  }\href {\doibase 10.1016/j.dark.2019.100411} {\bibfield  {journal} {\bibinfo
  {journal} {Physics of the Dark Universe}\ }\textbf {\bibinfo {volume} {27}},\
  \bibinfo {eid} {100411} (\bibinfo {year} {2020})},\ \Eprint
  {http://arxiv.org/abs/1907.08714} {arXiv:1907.08714 [gr-qc]} \BibitemShut
  {NoStop}%
\bibitem [{\citenamefont {{Khoury}}\ and\ \citenamefont
  {{Weltman}}(2004)}]{2004PhRvD..69d4026K}%
  \BibitemOpen
  \bibfield  {author} {\bibinfo {author} {\bibfnamefont {J.}~\bibnamefont
  {{Khoury}}}\ and\ \bibinfo {author} {\bibfnamefont {A.}~\bibnamefont
  {{Weltman}}},\ }\href {\doibase 10.1103/PhysRevD.69.044026} {\bibfield
  {journal} {\bibinfo  {journal} {\prd}\ }\textbf {\bibinfo {volume} {69}},\
  \bibinfo {eid} {044026} (\bibinfo {year} {2004})},\ \Eprint
  {http://arxiv.org/abs/astro-ph/0309411} {arXiv:astro-ph/0309411 [astro-ph]}
  \BibitemShut {NoStop}%
\bibitem [{\citenamefont {van~der Waals}(ates)}]{vdw}%
  \BibitemOpen
  \bibfield  {author} {\bibinfo {author} {\bibfnamefont {J.~D.}\ \bibnamefont
  {van~der Waals}},\ }\emph {\bibinfo {title} {1873}},\ \href@noop {} {Ph.D.
  thesis},\ \bibinfo  {school} {Universiteit Leiden} (\bibinfo {year} {Over de
  continuiteit van den gas- en vloeistoftoestand (On the Continuity of the
  Gaseous and Liquid States)})\BibitemShut {NoStop}%
\bibitem [{\citenamefont {Peralta}\ and\ \citenamefont {Jor{\'{a}
  }s}(2020)}]{Peralta_2020}%
  \BibitemOpen
  \bibfield  {author} {\bibinfo {author} {\bibfnamefont {C.}~\bibnamefont
  {Peralta}}\ and\ \bibinfo {author} {\bibfnamefont {S.}~\bibnamefont
  {Jor{\'{a} }s}},\ }\href {\doibase 10.1088/1475-7516/2020/06/053} {\bibfield
  {journal} {\bibinfo  {journal} {Journal of Cosmology and Astroparticle
  Physics}\ }\textbf {\bibinfo {volume} {2020}},\ \bibinfo {pages} {053}
  (\bibinfo {year} {2020})}\BibitemShut {NoStop}%
\bibitem [{\citenamefont {{Magnano}}\ and\ \citenamefont
  {{Soko{\l}owski}}(1994)}]{1994PhRvD..50.5039M}%
  \BibitemOpen
  \bibfield  {author} {\bibinfo {author} {\bibfnamefont {G.}~\bibnamefont
  {{Magnano}}}\ and\ \bibinfo {author} {\bibfnamefont {L.~M.}\ \bibnamefont
  {{Soko{\l}owski}}},\ }\href {\doibase 10.1103/PhysRevD.50.5039} {\bibfield
  {journal} {\bibinfo  {journal} {\prd}\ }\textbf {\bibinfo {volume} {50}},\
  \bibinfo {pages} {5039} (\bibinfo {year} {1994})},\ \Eprint
  {http://arxiv.org/abs/gr-qc/9312008} {arXiv:gr-qc/9312008 [gr-qc]}
  \BibitemShut {NoStop}%
\end{thebibliography}%

%

\end{document}